\documentclass{mn2e}
\input epsf.sty

\def\rtr{r_{\rm tr}}
\def\Rg{r_{\rm g}}
\def\Ns{N_{\rm s}}
\def\lh{l_{\rm h}}
\def\lhmax{l_{\rm h, max}}
\def\lsoft{l_{\rm s}}
\def\vff{v_{\rm ff}}
\def\trise{t_{\rm r}}
\def\tdelay{t_{\rm delay}}
\def\ttrav{t_{\rm trav}}
\def\rms{r_{\rm ms}}
\def\rout{r_{\rm out}}
\def\taues{\tau_{\rm es}}
\def\tauhot{\tau_{\rm hot}}
\def\Rflare{R_{\rm flare}}
\def\Fillum{F_{\rm irr}}
\def\sigmaT{\sigma_{\rm T}}
\def\me{m_{\rm e}}
\def\kT{k T_{\rm e}}
\def\MSun{{\rm M}_{\odot}}
\def\kapabs{\kappa_{\rm abs}}
\def\kapes{\kappa_{\rm es}}
\def\Sref{S_{\rm refl}}
\def\Sprim{S_{\rm prim}}
\def\Ka{K$\alpha$\ }

\title[A propagation model of X--ray variability]
{Simulations of X--ray spectral/timing properties in a propagation model 
of variability of accreting black holes}

\author[P. T. \.{Z}ycki]{Piotr T. \.{Z}ycki\thanks{e-mail: ptz@camk.edu.pl} \\
    Nicolaus Copernicus Astronomical Center, Bartycka 18, 00-716 Warsaw,
Poland}

\date{23 Nov 2002}

\begin{document}
\label{firstpage}

\maketitle

\begin{abstract}

A phenomenological model of X--ray variability of accreting black holes is 
considered, where the variable emission is attributed to multiple active 
regions/perturbations moving radially towards the central black hole.
The hard X--rays are produced by inverse Compton upscattering
of soft photons coming from reprocessing/thermalization of the same hard 
X--rays.
The heating rate of the Comptonizing plasma is assumed to scale with 
the rate of dissipation of gravitational energy while the supply of soft 
photons is assumed to diminish towards the center. Two scenarios are
considered: (1) an inner hot flow with outer truncated standard accretion disc 
and (2) an accretion disc with an active corona and a thick hot ionized skin.  
A variant of the model is also considered, which is compatible with 
the currently discussed multi-Lorentzian description of power spectral 
densities of X--ray lightcurves.

In the inner hot flow scenario the model can reproduce the observed Fourier 
frequency resolved spectra observed in X--ray binaries, 
in particular the properties
of the reprocessed component as functions of Fourier frequency.
In the accretion disc with ionized skin scenario the reduction of soft 
photons due to the ionized skin  is insufficient 
to produce the observed characteristics.

\end{abstract}


\section{Introduction}

X--ray emission from accreting compact objects carries information
about geometry and physical conditions of the accreting plasma. Spectral
and timing analyses of the data from sources in low/hard state have
enabled formulating a number of phenomenological scenarios for accretion
onto black holes. The scenarios belong to two main groups: (1) models
invoking standard optically thick accretion disc with an active corona and
(2) models assuming that the optically thick disc is truncated
at a radius larger than the last stable orbit, and is replaced
by inner hot, optically thin flow (see Done 2002, Czerny 2002 for reviews). 
Models
of both groups have some physical foundations. Magnetic dynamos in the disc
may 
give rise to the active corona, as suggested by recent magneto-hydrodynamical
simulations (Miller \& Stone 2000), with plasma outflow and/or hot
ionized disc skin responsible for the observed correlations
between spectral parameters: the power law slope and amplitude of
the reflected component (Zdziarski, Lubi\'{n}ski \& Smith 1999;
Gilfanov, Churazov \&  Revnivtsev 2000);
On the other hand, disappearance of the cold disc due to evaporation
may create the inner hot flow (R\'{o}\.{z}a\'{n}ska \& Czerny 2000).
Both classes of models can explain time average energy spectra,
as well as correlations between spectral parameters
(see discussion of models and more references in \.{Z}ycki 2002;
 hereafter Z02).

Specific models of X--ray variability were constructed for both 
geometries (see  Poutanen 2001 for review). In the active corona geometry
the model of magnetic flare avalanches of Poutanen \& Fabian (1999; 
hereafter PF99) 
is very successful in reproducing
a number of observed properties (e.g.\ power spectra, hard X--ray time lags).
However, its simple extension to accretion disc with the ionized skin
fails to reproduce the  Fourier-frequency resolved spectra 
(Revnivtsev, Gilfanov \& Churazov 1999, 2001) observed from black hole binaries
(Z02). In the inner hot flow geometry the drifting blob model 
(B\"{o}ttcher \& Liang 1999) reproduces certain observed characteristics.
However,  Maccarone, Coppi \& Poutanen (2000) argue that flares in hard
(Comptonized) X-rays are the primary driver of variability, the soft
(thermal) emission being merely reprocessed hard X-rays (see also 
Malzac \& Jourdain  2000).

The observed logarithmic energy dependence of the hard X--ray time lags
was shown to be compatible with a simple propagation model,
where X--ray emitting perturbations move radially towards the central
black hole, the emitted spectrum hardening as the perturbations
approach the center (Miyamoto \& Kitamoto 1989; Nowak et al.\ 1999;
Kotov, Churazov \& Gilfanov 2001). 
Moreover, the observed characteristic 'dip'
in the energy dependence of time lags at the energy of the Fe \Ka line
can be explained in the propagation model, if the line emission from inner
disc is suppressed (Kotov et al.\ 2001). 

Therefore in this paper we consider a phenomenological propagation model,
where the variable X--ray emission is attributed to active 
regions and/or perturbations moving radially towards the central compact
object. X--rays are assumed to be produced by  inverse Compton
upscattering of soft photons coming from a cool, optically thick disc.
X--ray luminosity is assumed to increase as the emission progresses,
thus creating a flare of radiation. The Comptonized spectrum is assumed
to evolve from softer to harder during the flare due to diminishing supply 
of seed photons. This may be due to either the optically thick disc
being absent at small radii, or the thickness of the ionized disc skin
increasing towards the center. From the simulated event files a number of 
statistics will be computed, which may be compared to available data. In 
particular, the Fourier-frequency resolved spectral properties 
 are studied: both the continuum slope and
Fe \Ka line strength as functions of Fourier frequency, $f$, will be computed.
The observed Fourier-frequency resolved spectra, calculated from the 
{\it RXTE\/} data of black hole binaries in the low/hard state, 
show two important characteristics:
the continuum spectra get harder with increasing $f$, while the amplitude
of reflection decreases with $f$ (Revnivtsev et al.\ 1999, 2001).

We emphasize that the main feature of the considered models is the radial 
propagation of the 
X-ray emitting structures, with the related spectral evolution. 
This is meant to be compared with the models assuming radially localized
flares (e.g.\ the model of PF99). However, we do not 
hypothesize here about possible correlations between flares
(although the idea of flare avalanches will be implemented in
Sec~\ref{sec:aval}), hence this
model is not meant to explain in detail the X--ray power spectra and higher
order statistics of the variability (Maccarone \& Coppi 2002). 

Plan of the paper is as follows: the model is described  in 
Sec.~\ref{sec:model}, results 
for the hot inner flow geometry are presented in Sec.~\ref{sec:results},
while results for the case of an accretion disc with hot ionized skin
are presented in Sec.~\ref{sec:hotskin}. In Sec.~\ref{sec:multilor}
a variant of the model is considered, which is compatible with
the multi-Lorentzian description of power spectra (Nowak 2000).

\section{The Model}
\label{sec:model}

\begin{figure}
\epsfxsize = 0.5\textwidth
\epsfbox[20 420 600 720]{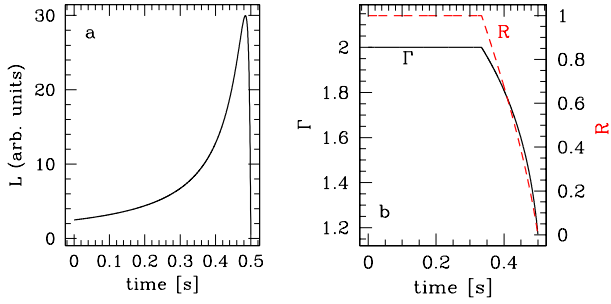}
\caption{Time variations of parameters during a flare in the truncated disc
scenario. (a) luminosity of a flare, (b)
 spectral index, $\Gamma$, and amplitude of reflection, $R$. 
Parameters: $\rout=60\,\Rg$, $\beta=0.02$ so that the flare duration is
$\ttrav=0.5$ sec, $\rtr=30\,\Rg$.
\label{fig:flare}
}
\end{figure}

The X--ray emitting structures are assumed to originate at a
certain radius $\rout$, which will be assumed $\sim 50\,\Rg$
($\Rg\equiv GM/c^2$; precise value of $\rout$ has little influence on 
our results, since the contribution
to emission from large distances is insignificant). They move towards
the center at a fraction of the free-fall speed,
\begin{equation}
\label{equ:motion}
 v = \beta \vff = \beta \sqrt{ {2 G M \over r}},
\end{equation}
where $\beta$ is a model parameter. Since the heating of the plasma must 
ultimately come from the dissipation of the gravitational energy, we assume
that the plasma heating rate scales as the dissipation rate of the latter for
a disc-like accretion,
\begin{equation}
 \label{equ:lh}
 \lh(t) \propto r(t)^{-2}\,b[r(t)],
\end{equation}
where the exponent $-2$ corresponds to dissipation per ring of matter.
Here $l$ is the compactness parameter, 
$l\equiv {L\over \Rflare}{\sigmaT\over \me c^3}$
($\Rflare$ is the characteristic radius of the structure),
and $b(r)$ is the boundary term which is assumed to have its Newtonian
form $b(r) = 1-\sqrt{6\Rg/r}$ (Shakura \& Sunyaev 1973). 
Other prescriptions for $\lh(r)$ could be used here, since the relation 
between
the overall disc accretion flow and the motion of structures is obviously
uncertain\footnote
{One subtle geometrical effect here is the radial dependence of the size
of the emiter, $\Rflare$.  
Actually, it is the luminosity from each emitter that 
can be expected to scale as $r^{-2}$. Then, eq.~\ref{equ:lh} is equivalent
to an assumption $\Rflare={\rm const}$. If $\Rflare$ varies with $r$, radial 
dependencies
of $L$ and $\lh$ will be different, but the observed quantity is $L$
and this should scale as $r^{-2}$. Precise values of  $\lh(r)$
is not important in our paper because of approximations used to compute
the spectra (eq.~\ref{equ:lsoft}). In particular, pair production
processes are neglected. This problem will be more apparent
in Section~\ref{sec:hotskin}.
}. 
The heating rate
is normalized to have an assumed maximum value, $\lhmax$. Solving 
the equation of motion (Eq.~\ref{equ:motion}) gives 
\begin{equation}
r(t) = (\rout^{3/2}-A t)^{2/3},
\end{equation}
where 
$A={3\over 2}{ \sqrt{2} \beta c \over \Rg}$, and the radial positions
are expressed in units of $\Rg$. The duration of a flare is
$\ttrav=(\rout^{3/2}-6^{3/2})/A$, since the final radial position is 
assumed $6\,\Rg$. The resulting
flare, $\lh(t)$, is characterized by a slow rise and a sudden decrease
(Fig.~\ref{fig:flare}a)
in agreement with the constraints derived by Maccarone et al.\ (2000). 

The soft photons are assumed to
come from reprocessing and thermalization of the hard X--rays, so that
the feedback loop is realized as needed to explain the $R$--$\Gamma$
correlation (Zdziarski et al.\ 1999; Gilfanov et al.\ 2000). 
Two geometrical scenarios may be considered here:
\begin{enumerate}
\item  an inner hot flow
partially overlapping a cold, optically thick disc disrupted at certain 
radius, $\rtr$, (see e.g.\ Poutanen, Krolik \& Ryde, 1997, for arguments for 
the overlap). The luminosity of soft photons crossing an active region 
is therefore parametrized as
\begin{equation}
 \label{equ:lsoft}
  \lsoft(t) = \lh(t) \times \Ns \left\{
 \begin{array}{cc}
             1                         & \mbox{for } r \ge \rtr \\
    \left({r \over \rtr} \right)^2  & \mbox{for } r < \rtr,
 \end{array}
 \right.
\end{equation}
where $\Ns \approx 0.5$, 
\item a standard optically thick accretion disc with an 
active corona and a hot ionized skin on top of the disc (Nayakshin \& Dove
2001). The flux
of soft photons crossing an active region is computed using the
method of \.{Z}ycki \& R\'{o}\.{z}a\'{n}ska (2001) and Z02,
for an assumed  height of the active region above the disc,
as it travels towards the center.
\end{enumerate}

It should be noted here that for a truncated disc, wholly replaced by
a hot flow, and for a compact emission region, the resulting dependence
$\lsoft(r)/\lh(r)$ is much steeper than the assumed $r^{2}$. That is,
once the active region is within the central hole, it can intercept only 
a tiny fraction of the thermalized emission, if its size is comparable
or smaller than the transition radius. This would produce much harder
spectra than observed, $\Gamma < 1$. Therefore we choose here
a weaker dependence, although the particular value of the exponent
is chosen rather
arbitrarily and no attempt is made here to tune it to the
observed data.
The slow dependence may be interpreted as the optically thick plasma
slowly disappearing (disc disrupted into small clouds; 
covering factor gradually decreasing from 1 to 0) with decreasing $r$,
or the emission coming from perturbations propagating in the
entire volume of the hot plasma.

The emitting structures are assumed to originate at $\rout\ge \rtr$ at a mean
rate of $\lambda$ per second. Time intervals between the launch of
successive flares are generated from the $\lambda \exp(-\lambda t)$ 
distribution, as appropriate for a Poissonian process.
Each flare is followed until it reaches the marginally
stable orbit at $\rms = 6\,\Rg$. At each time step the primary Comptonized
component and its reprocessed component are computed. The truncation radius
is assumed $\rtr=30\,\Rg$, so that the time average spectral slope of the
Comptonized continuum is $\Gamma=1.6$--1.7.

The  Comptonized spectrum is computed using the code
{\sc thComp} (Zdziarski, Johnson \& Magdziarz 1996), solving the Kompaneets
equation. Computations are parametrized by the photon spectral index,
which we compute from $\Gamma = 2.33(\lh/\lsoft)^{-1/6}$ (Beloborodov
1999a,b), and electron temperature 
$\kT/(\me c^2)$ computed using formulae from Beloborodov (1999b).
Plasma optical depth is assumed $\taues = 1.8$ (PF99). 

The reprocessing medium is 
assumed to be 'cold', (only hydrogen and helium ionized) with elements
abundances of Morrison \& McCammon (1983). The spectrum of the Compton
reflected continuum is computed from the simple formula of
Lightman \& White (1988),
\begin{equation} 
 \label{equ:lwrefl} 
  \Sref(E) = {1-\epsilon \over 1+\epsilon} \Sprim(E), \quad\quad 
      \epsilon = \sqrt{ {\kapabs \over \kapabs + \kapes}}, 
\end{equation} 
where $\kapabs(E)$ and $\kapes(E)$ are the photo-absorption and electron 
scattering opacities, respectively. The above formula is multiplied by a 
simple exponential cutoff to mimic the high energy cutoff. This is sufficient
for our purposes since only the spectra up to $\approx 30$ keV will be
considered. The Fe \Ka line is added to the reflected continuum,
with equivalent width (EW) as a function of $\Gamma$, following computations
of \.{Z}ycki \& Czerny (1994; see also Z02). Importantly,
the instantaneous amplitude of the reprocessed component is assumed to 
depend on radius of emission,
\begin{equation}
 \label{equ:refl}
  R(r) = \left\{
 \begin{array}{cc}
             1                         & \mbox{for } r > \rtr \\
     {r-\rms \over \rtr-\rms}   & \mbox{for } r < \rtr,
 \end{array}
 \right.
\end{equation}
i.e.\ it linearly decreases from 1 at $r=\rtr$ to 0 at $r=\rms$. Again,
the particular form of the function $R(r)$ for $r<\rtr$ is chosen
rather arbitrarily, the general condition being that $R(r)$ decreases
inwards. 
Time evolution of $\Gamma$ and $R$ during a flare is plotted in
Fig~\ref{fig:flare}b.

Thus created sequence of spectra is subject to standard analysis in 
the time and Fourier domains (see e.g.\ van der Klis 1995; 
Nowak et al.\ 1999; Poutanen 2001).
In particular, the Fourier frequency resolved ($f$-resolved hereafter)
spectra are computed according to the
prescription given by Revnivtsev et al.\ (1999; see also Z02). 
The normalized power spectral density (PSD) is computed as
\begin{equation}
 P_j = 2 {T \over \bar{C}^2} |C_j|^2, \quad\quad j=0,\dots,N-1
\end{equation}
\begin{equation}
 C_j = \sum_{k=0}^{N-1}\,c_k\,e^{2\pi i f_j t_k},
\end{equation}
where the discrete frequencies $f_j\equiv j/T$ for $j=-N/2,\dots, N/2$;
$T$ is the total light curve time, $\bar{C}$ is the mean count rate,
while $c_k$ are the number of counts in $k$-th time bin. The $f$-resolved
spectrum at energy $E_i$ and Fourier frequency $f_j$ is then defined as
\begin{equation}
 \label{equ:ffres}
S(E_i, f_j) = \bar{C}_i\sqrt{P_i(f_j)\,\Delta f_j} = 
 \sqrt{ {2 |C_{i,j}|^2 \over T} \Delta f_j}.
\end{equation}

All computations are done assuming the central black hole mass $M=10\,\MSun$.

\section{Results}
 \label{sec:results}

\begin{figure*}
 \epsfxsize = 0.95\textwidth
 \epsfbox[18 520 600 710]{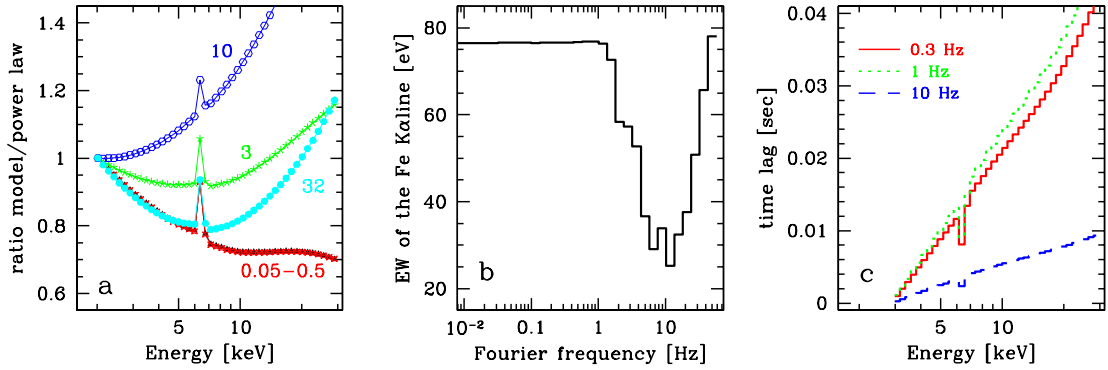}
 \caption{Results of simulations for a single value of active regions 
 velocity. Parameters: $\beta=0.02$, $\rout=60\,\Rg$, so that the duration
  of a flare is $\ttrav\approx 0.5$ s (Sec.~\ref{sec:single}). 
 (a) The $f$-resolved spectra, divided by
 a power law, $N_E\propto E^{-1.6}$, and normalized to 1 at 2 keV
 (labels 
  are middle frequencies in $\Delta f_j$ in Hz as in equation~\ref{equ:ffres}) 
 The spectra harden initially with $f$ up to 
 $f\sim 10$ Hz, then soften, with  the corresponding increase of the
 reflection amplitude. (b) Equivalent width  of the Fe \Ka line
 showing dependence on $f$ matching that of the $f$-spectra.
 (c) Time lags as functions of energy, for three values of
  Fourier frequency (Kotov et al.\ 2001; also fig.~12 in Nowak et al.\ 1999).
\label{fig:singlevel}}
\end{figure*}

\begin{figure*}
 \epsfxsize = 0.95 \textwidth
 \epsfbox[18 520 600 710]{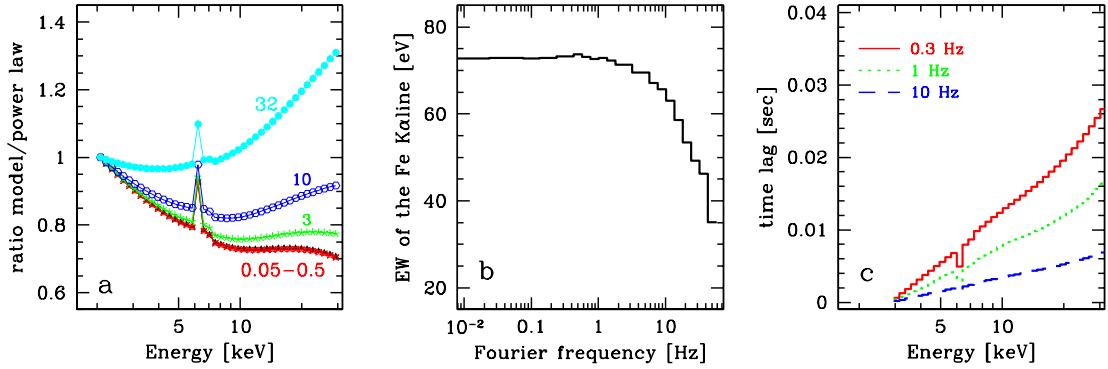}
 \caption{Results of simulations for avalanches of flares. Curves are 
 described in the same way as in Fig.~\ref{fig:singlevel}. Duration of the
 flares covers the range 0.03 s to 1.5 s ($\beta=0.35$--$0.007$, respectively).
 The $f$-resolved
 spectra and EW of \Ka line are now monotonic functions of Fourier
 frequency $f$, similarly to the observed dependencies 
 (Revnivtsev et al.\ 1999). The $\delta\tau(E)$
 dependence, panel (c), reproduces the observed dependencies well
 (compare with fig.~12 in Nowak et al.\ 1999).
\label{fig:avalanches}}
\end{figure*}

\begin{figure*}
 \epsfxsize = 0.95 \textwidth
 \epsfbox[18 520 600 710]{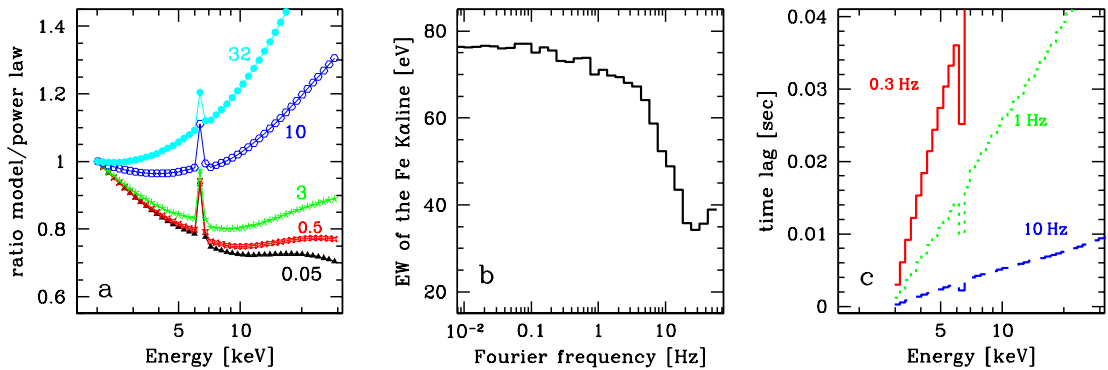}
 \caption{Results of simulations for a simple distribution of velocities of
 emission regions (no correlations between flares, i.e.\ no avalanches).
 Curves are  described in the same way as in Fig.~\ref{fig:singlevel}. 
 The $f$-resolved  spectra and EW of \Ka line are very similar to the
 avalanche case (Fig.~\ref{fig:avalanches}), but the time lags 
 $\delta\tau(E)$ are much longer and  longer than those  observed in Cyg X-1.
 Longer individual flares are necessary to generate variability power at 
 frequencies  down to 0.1 Hz.
\label{fig:distribvel}}
\end{figure*}

\subsection{The case of a single value of velocity of the active regions}
\label{sec:single}

In the simple case of a single value of $\beta$ we do not expect
the observed PSD to be reproduced, since the $f^{-1}$ behaviour generally
requires a distribution of parameters and/or correlations between flares
(Lehto 1989, Lochner, Swank \& Szymkowiak  1991).
It is nevertheless instructive
to study other observables from the model.  The velocity proportionality 
coefficient was adopted $\beta=0.02$ and $\rout=60\,\Rg$,
so that the duration of a flare is $\ttrav\approx 0.5$ s, and the PSD 
peaks at 0.5--1 Hz.
Fig.~\ref{fig:singlevel} shows the $f$-resolved spectra and 
time lags as functions of energy.
The model predictions do not match well the observed trends.
The 
time lags as a function of energy are independent of Fourier frequency,
at least for $f\le 1$ Hz, that is below the peak frequency of PSD.
This contrasts the observed $\delta\tau(E)$, which
shows longer lags at a given $E$, for lower $f$ (Nowak et al.\ 1999).
The logarithmic dependence of energy is true for any flare profile,
if the change of spectral index during the flare, $\Delta \Gamma$, 
is small compared
to time average $\Gamma$ (Kotov et al.\ 2001). In the considered case
$\Delta\Gamma\approx 0.8$, so it is not small compared to 
$\langle \Gamma\rangle\approx 1.7$, but the logarithmic dependence
$\delta\tau(E) \propto \log E$ is still obeyed, at least for higher 
$f\ge 1$ Hz, and energies not exceeding $\sim 30$ keV. This is consistent
with analytical solution, which can be obtained for a simple exponential
profile and a linear dependence of $\Gamma(t)$.
Assuming
\begin{equation}
 S(E,t) = \exp\left(t/\trise\right) E^{-\Gamma_f + a t},\ \ \ \ t<0,
\end{equation}
where $\Gamma_f$ is the spectral index at the peak of a flare, at $t=0$,
the Fourier transform can be computed,
\begin{equation}
 \hat{S}(E,f) = E^{-\Gamma_f}{ \trise \over 1-i \trise 2 \pi f + a \trise \ln E}.
\end{equation}
The phase lags are 
\begin{equation}
 \phi (f) \equiv \arg\left[ \hat{S}(E_1,f) \hat{S}^{*}(E_2,f)\right],
\end{equation}
where
\begin{eqnarray}
 \tan \phi(f) = &\nonumber \\ 
 & {-a \trise^2 \ln(E_1/E_2) 2\pi f \over 
 1+ a\trise\ln(E_1 E_2) + a^2 \trise^2    \ln E_1 \ln E_2 + \trise^2 (2\pi f)^2}.
\end{eqnarray}
From this it can be seen that for small $\phi$, the time lags $\delta\tau(E)$
are independent of $f$, $\delta\tau = \phi/(2\pi f) \approx \tan\phi/(2\pi f)
\approx {\rm const}$.

The dip in $\delta\tau(E)$ at the energy of the Fe \Ka line, 
seen in the real data, is  reproduced by the model, confirming the result 
of Kotov et al.\ (2001).

The observed spectral slope of the comptonized continuum and 
the equivalent width of the Fe \Ka line are monotonic functions 
of $f$, at least above $\approx 0.5$ Hz (Revnivtsev et al.\ 1999).
In contrast to these, the model
$f$-resolved spectra show complex dependence on $f$. In particular
the EW of the \Ka line first decreases above $\approx 1$ Hz,
then increases again above $\approx 10$ Hz. This means that
in the Fourier decomposition of the flare profile the highest frequencies
form the first part of the rising profile, which, in real space,
is formed above the accretion disc (i.e.\ $\rtr < r < \rout$). The spectra
from that part are soft, and the amplitude of reflection
is close to 1. However, we note that the $f$-resolved spectra are dependent on
the radial profile of the heating function (Eq.~\ref{equ:lh}). Experimenting
with the $\lh(r)$ function we find that a more
peaked $\lh(r)$  (e.g.\ $\propto r^{-\gamma}$, $\gamma\ge 3$) produces
the $f$-resolved spectra monotonically hardening with $f$ and
EW$(f)$ monotonically decreasing.

\subsection{Distribution of parameters}
\label{sec:aval}

The full model obviously has to explain all observable characteristics
of the X--ray emission. The complex but universal PSD of X--ray emission
in low/hard state points out towards a universal mechanism of generating
the complexity. One such mechanism was suggested by Stern \& Svensson
(1996) in the context of gamma-ray bursts and implemented by 
PF99 for accreting black holes. Their idea is that
each flare can stimulate one or more flares to produce an 'avalanche' of 
many flares. Parameters of avalanches can be adjusted so that the power 
spectrum can have the observed slope of $-1$ (i.e.\ $P(f) \propto f^{-1}$)
in a given frequency band.

The idea of avalanches of flares may be easily implemented in the present 
model. Different time-scales of individual flares correspond to 
different coefficients
$\beta$, and the stimulated flares are assumed to originate from the same 
radius as the stimulating ones. Specifically, we assume that $\beta$ follows
a power law distribution $P(\beta) \propto \beta^k$ between $\beta_1$ and 
$\beta_2$. There are $\lambda$ spontaneous flares per second, and each 
flare has a probability $\mu$ of stimulating a secondary flare, the
latter being delayed by $\tdelay=\alpha \,\ttrav/7$, 
where  $\ttrav$ is the duration of the stimulating 
flare, while $\alpha$ is drawn from a Poisson distribution of
mean $\alpha_0$ (the factor 7 approximately translates $\ttrav$ to flare 
time scale used by PF99; see there for detailed description of the model).
The adopted values of the parameters are such that they give good description
of PSD from black hole binaries: $k=0.5$, $\beta_1 = 0.007$,
$\beta_2=0.35$, $\lambda=90$, $\mu=0.7$, $\alpha_0=5$. The perturbations
start from $\rout=60\,\Rg$, so the duration of the flares cover the range
from 0.03 to 1.5 sec.

Results for this model are presented in 
Figure~\ref{fig:avalanches}. They match the observed dependencies reasonably
well. Time lags as functions of energy are in good quantitative agreement 
with the observed values in the whole range of computed $f$ (compare
with Cyg X-1 data in Nowak et al.\ 1999).
Interestingly, by superposing
flares of different durations, the rising trend in EW$(f)$ for $f>10$ Hz
(see Fig.~\ref{fig:singlevel}) 
was compensated for, and the equivalent width is now monotonically
decreasing with $f$. Correspondingly, the continuum hardens monotonically
with $f$. Both trends are as observed in black hole binaries
(Revnivtsev et al.\ 1999, 2001).
This means that not only are higher frequencies
formed from shorter flares, but within the flares the dominant contribution
to a given $f$ moves towards the peak of the flare, as $f$ increases.

For comparison, Fig.~\ref{fig:distribvel} shows  results obtained
assuming a simple distribution of velocities of the emission structures
(no correlations between flares). 
Coefficients $\beta$ was drawn from a uniform distribution between
$\beta_1 = 0.003$ and $\beta_2 = 0.1$ (flare durations 0.1 s to 3.5 s). 
Importantly, the lower value
$\beta_1$ is now much smaller than that in the avalanche prescription,
to give flat $P(f)$ down to $\approx 0.1$ Hz. As a result, the time lags
are now longer than those for avalanches, clearly longer than those
observed.

\subsection{Accretion disc with hot ionized skin}
 \label{sec:hotskin}

The presented propagation model  can also be formulated in the geometry of 
an accretion
disc extending down to the last stable orbit, with the X--ray emitting
structures 
moving above the disc towards the center. As a result of irradiation by
hard X--rays, the hot ionized skin will form on the disc
(R\'{o}\.{z}a\'{n}ska \& Czerny 1996; Nayakshin \& Dove 2001). Making the same
assumption about increasing luminosity of the moving flare as in 
Section~\ref{sec:model} (equation~\ref{equ:lh}), the result is
obtained that the thickness of the hot skin, $\tauhot$, 
increases towards the center
(Nayakshin 2000; \.{Z}ycki \& R\'{o}\.{z}a\'{n}ska 2001). 
This reduces the effectiveness of reprocessing/thermalization as the
flare progresses. This in turn leads to diminishing supply of soft photons for 
Comptonization and hardening of the continuum as a consequence. Simultaneously,
the relative amplitude of the ``cold'' reprocessed component decreases, since
fewer photons penetrate to the cold disc beneath the hot skin
(e.g.\ Done \& Nayakshin 2001; \.{Z}ycki \& R\'{o}\.{z}a\'{n}ska 2001). 
Qualitatively, the situation is thus analogous to the disrupted disc
scenario, and could reproduce observed trends of $\Gamma$ and $R$
with Fourier frequency.  In this Section  a quantitative analysis of the
model is performed.

\begin{figure}
\epsfxsize = 0.5\textwidth
\epsfbox[20 420 600 720]{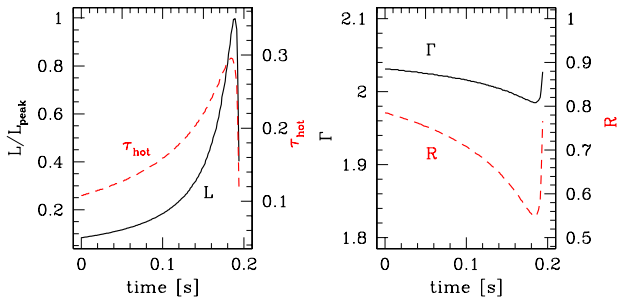}
\caption{Time variations of parameters during a flare in the hot skin
scenario (Sec.~\ref{sec:hotskin}). Left panel: luminosity of a flare (solid 
curve) and the thickness of the hot skin (dashed curve), 
right panel: spectral index,
$\Gamma$, and amplitude of reflection, $R$. Note the relatively small
range of $\Gamma$ and $R$, due to small variation of $\tauhot$.
 The spectral evolution during a flare is
{\em insufficient\/} to produce correct hard X--ray time lags or $f$-resolved
spectra.
\label{fig:skinflar}
}
\end{figure}

Basic model element is the spectral evolution during a single flare.
Here the radial dependence of the disc illuminating flux,
$\Fillum(r)$, is crucial since it determines the $\tauhot(r)$ dependence.
The illuminating flux can be written as $\Fillum = (\eta/\pi) 
L/(\Rflare^2 + 3 H^2)$, 
where $\eta\approx 0.5$ is the anisotropy factor, $\Rflare$ is the radius
of a flare and $H$ is its height
above the disc (see PF99). Since, as
argued above (equation~\ref{equ:lh}), $L(r) \propto r^{-2}$, we would obtain
$\Fillum(r) \propto r^{-2}$ if $\Rflare={\rm const}$ and $H={\rm const}$.
However, for $\tauhot$ to have a significant radial dependence, $\Fillum$
must be a stronger function of $r$. For example, for illumination due to
local dissipation, $\Fillum(r) \propto r^{-3}$. To achieve this, we assume
here $\Rflare(r)\propto H(r) \propto r^{1/2}$. That is, the flare shrinks and
its height decreases as it approaches the center. The condition
$\Rflare \propto H$ is required in order to reproduce the $\Gamma$--$R$
correlation, since it ensures that the geometry of the flare is constant
and the only varying factor is the thickness of the hot skin.
The proportionality coefficient in the $\Rflare(r)=R_0 r^{1/2}$ 
relation can be computed
if the maximum heating compactness, total source luminosity
and flare occurence rate are known (see eq.~7 in Z02).

Energy spectra during a flare are computed as in Z02. 
The thickness of the hot skin, $\tauhot$, is computed as a function of 
$\Fillum$ and $r$ using the fitting formula (eq.~9) from that paper, while
the amount of soft photons from thermalization re-entering an active region 
and the effective amplitude of reflection were computed from
eqs.~3, 11 and 10  in Z02, respectively, following PF99 and Poutanen (2002).
The feedback parameter $D_0 = 0.5$ (eq.~3 in Z02 and PF99)\footnote{
We note that the parametrization adopted by PF99 hides the fact that
the required $D_0$ is actually incompatible with an emission region whose
horizontal and vertical extent are comparable, if the emission
is {\it isotropic\/}. According to PF99 for 
$D_0=0.5$ one half of 
the emitted X--rays return to the emission region if $H=0$. This in reality 
implies the horizontal extent much larger than vertical one 
(slab-like geometry).}.
The height $H$ is assumed to be 0, so that in the absence of the
hot skin $\Gamma\approx 2$.
It should be emphasized that in the present scenario the time/radial
dependencies of $\lh/\lsoft$ and $R$ can be predicted from physical
considerations. Thus, eq.~\ref{equ:lsoft} and eq.~\ref{equ:refl} are
{\em not\/} used here.

An example of a flare computed in this model is shown in 
Fig~\ref{fig:skinflar}. The spectrum does evolve from soft to hard, but
the change of $\Gamma$ is very small, clearly insufficient to
produce any difference in $f$-resolved spectra.
A sequence of spectra was simulated, assuming a distribution of 
velocities of active regions and other parameters 
as in pulse avalanche prescription (Sec.~\ref{sec:aval}).
Computing the $f$-resolved spectra we confirm that they all have the same
slope, $\Gamma\approx 2$, while the EW of the \Ka line is approximately
constant at $\approx 55$ eV. Thus, the hot skin disc model in the simple
version considered here is not able to reproduce the $f$-resolved spectra.

The irradiation flux would have to be a very strong function of $r$
for the hot skin thickness to increase enough to produce the required
hard spectrum ($\Gamma\sim 1.6$--$1.7$ at the peak of the flare).
This would mean a significant decrease of the size of the emitting region as 
it approaches the center, if the luminosity $L(r) \propto r^{-2}$ dependence 
is kept. That in turn would increase the compactness parameter, $\lh$,
as well as the temperature of the soft thermalized radiation.
The former cannot be uniquely constrained by the data, if the plasma
is thermal (Macio{\l}ek-Nied\'{z}wiecki, Zdziarski \& Coppi 1995), but
the latter is observed to be no higher than $T_0 \sim 0.3$ keV
(e.g.\ Di Salvo et al.\ 2001). It remains to be seen whether a physically
plausible model can be constructed fulfilling all the constraints.

\section{A model compatible with multi-Lorentzian decomposition of power
spectra}
\label{sec:multilor}

X-ray power spectra from accreting compact objects can be interpreted in terms
of a (small) number of relatively narrow Fourier components 
(Nowak 2000; Belloni, Psaltis \& van der Klis 2002).
Higher quality data from RXTE clearly reveal that PSD, which were previously
considered a roughly featureless power law with two or three breaks, 
consist of a number of peaks. This interpretation is strengthened by 
the correlations between frequencies of the  components 
(Belloni et al.\ 2002), 
reducing the complexity of broad band variability to possibly just one
parameter, at least within a given spectral state. The parameter seems
to be related to the geometry of accretion flow, as evidenced by
correlations between the characteristic frequencies and energy spectral
properties (spectral slope, amplitude of the reprocessed component; 
Revnivtsev et al.\ 2001; Gilfanov et al.\ 2000).
The obvious interpretation of the geometrical factor is the variable 
transition radius between the outer optically thick standard accretion 
disc and inner optically thin hot flow (Revnivtsev et al.\ 2001; 
Psaltis \& Norman 2002).

The narrow components of PSD are usually  modeled as Lorentz profiles.  
If the Lorentz profiles are taken 
seriously, the emission may be envisioned to consists of a collection of
exponentially damped (or forced i.e.\ rising) harmonic oscillators. 
Obviously, the oscillations may be reduced to simple exponential shots, 
if the quality factor of the Lorentzians, $Q\sim \Delta f/f_0 $
($f_0$ is the resonant frequency while $\Delta f$ is the half-width
of the Lorentzian), is low enough, $Q\le 1$. 
In the interpretation of Psaltis \& Norman (2002), the frequencies
would be connected to characteristic frequencies at the
disc truncation radius: orbital, periastron-precession and
nodal-precession  (Stella \& Vietri 1998).
However, it is the hard X-ray emission that is actually variable, and
the most likely reason for the variability is a variable energy dissipation
in the hot plasma, rather than variable input of soft photons
(Maccarone et al.\ 2000; Malzac \& Jourdain 2000), even though it is far from
obvious how to make a connection between the {\rm disc\/} frequencies
and the energy dissipation in the hot plasma. Furthermore, hard X-ray
time lags can be interpreted as a result of spectral evolution during
individual emission events (shots/flares in the usual shot noise models;
PF99; Kotov et al.\ 2001).

In this Section we consider a descriptive model compatible with the above
ideas. Adopting the form of the Lorentzian function as in Nowak (2000),
\begin{equation}
 P(f) = \pi^{-1} { N^2 Q f_0 \over f_0^2 + Q^2 (f-f_0)^2 }
\end{equation}
where $N$, $Q$ and $f_0$ are amplitude, quality factor and resonant frequency
of the oscillations, we simulate the shots of the following temporal profile:
\begin{equation}
 A(t) = A_0 \exp\left( {t \over \trise} \right)\, \cos(2\pi f_0 t + \phi_0),\ \ \ \ t<0.
\end{equation}
Here, the normalization $A_0=N Q^{1/2}$, 
the rise time $\trise = 2 Q/(2\pi f_0)$
and $\phi_0$ is a randomly chosen initial phase. Parameters of the Lorentzians
are taken from fits to Cyg X-1 0--4 keV light curves (Table 2 in Nowak 2000).
We ignore the lowest amplitude component QPO$_2$, which leaves us with 5
components. 

The crucial assumption we make here is that the shots correspond to 
perturbations traveling through the
plasma, from the truncation radius, $\rtr$, towards the center, at a speed
which may, but does not need to be, the same for all components.
In actual computations the speed is assumed constant in time, 
but assuming $v\propto\vff$ (as in previous Sections) would not change
our conclusions.  
The heating rate of the plasma is assumed to be proportional
to the energy of the oscillations, i.e. $\lh(t) \propto [A(t)]^2$, as in 
usual oscillatory motion (the square of a damped/forced oscillator is again
a damped/forced oscillator with a Lorentzian PSD). 
We emphasize that the heating is now {\em not\/} described by
eq.~\ref{equ:lh}, since it is determined by $A(t)$, but formulae
for $\lsoft(t)$ and $R(t)$ (eq.~\ref{equ:lsoft} and eq.~\ref{equ:refl})
are adopted.
Each of the 5 components is chosen with the same probability,
while the overall frequency of shots is adjusted to give the r.m.s.\ 
variability
$\approx 30$ percent. Computations of spectra and subsequent Fourier
analysis proceeds as in previous models.

In the particular case considered here the longest $\trise \approx 1.2$ sec
for the lowest $f_0$ component, and $\trise$ is progressively shorter for 
the higher-$f_0$ components.
The propagation speed, $v$, is chosen such that the travel time,
$\ttrav = (\rtr-r_{\rm f})/v$ is at least $5\trise$. This means that
a shot is launched from $\rtr$ at $t=-5\trise$, where $\trise$ is
the corresponding rise time-scale (the beginning of an exponential shot
is obviously not well defined). The final minimum radius, $r_{\rm f}$, 
reached at $t=0$, is determined by $v$.

It is easy to predict qualitatively the results for $f$-resolved spectra, 
if the propagation speed is the same for all components. 
The higher the $f_0$ (the shorter the $\trise$),
the shorter the distance traveled. Shorter distance means softer energy
spectrum, since spectra produced closer to the truncation radius are,
by construction, softer.
Thus the $f$-resolved spectra can be expected to become softer with increasing
$f$, opposite to those observed. Thus, obviously, a necessary condition for
this model to reproduce the observed $f$-spectra is to assume that the
speed of the perturbations increases with $f_0$. Then, 
the higher the frequency $f$, the farther from the cold disc the emission 
takes place, producing harder spectra. Adjusting the speeds 
accordingly we indeed obtain the required dependence of $f$-spectra
(including EW of the \Ka line) on Fourier frequency.

\section{Discussion}
\label{sec:discuss}

We have considered a propagation model of X--ray variability in
application to high energy emission from accreting compact objects. 
The motivation for our considerations comes from two directions:
(1) problems faced by the models of localized flares 
(PF99) in explaining the $f$-resolved spectra
(Z02), and (2) demonstration by Kotov et al.\ (2001)
that a propagation model with spectral evolution 
can naturally explain the hard X--ray time lags.
The model can be formulated in both considered geometries:
truncated standard disc with inner hot flow and a standard disc with
an active corona and the hot ionized skin. It should be emphasized,
however, that the main parameters entering the spectral computations,
$\lh/\lsoft(t)$ and $R(t)$ can be {\em predicted\/} only in the latter model, 
while they have to be assumed rather arbitrarily in the former. It is 
the greater predictive power of the hot skin scenario what enables to
find unique model predictions and demonstrate the problems which the
model faces. 

Similar computations of variability could be performed in the context
of the plasma  outflow model of Beloborodov (1999a,b). Here the main
parameter is the outflow speed. Qualitatively, the requirement for
the model to work is that the speed increases inwards. A good physical
model predicting the speed as a function of distance, X--ray flux
and other parameters is necessary, in order to make quantitative predictions.

Let us concentrate now on the geometry of an inner hot X--ray producing flow 
with an outer standard optically thick disc. 
This is one of the possibilities of the geometrical arrangement of
accreting multi-phase plasma. It naturally accounts for hard spectra
of black hole X--ray binaries, small amplitude of the reprocessed
component, and observed correlations between these spectral parameters
(Done 2002 and references therein).
Emission in the form of flares (shots) associated with active regions and/or
perturbations 
traveling in the hot plasma might then be responsible for observed time 
and spectral variability, considering that X--ray variability
is generally of stochastic character (Czerny \& Lehto 1997).
Such an extension of models of steady emission from hot flows
(e.g.\ Esin, McClintock \& Narayan 1997) is necessary, considering
the many observable characteristics related to variability. 
The situation is fully analogous to models invoking  active regions 
above an accretion disc (e.g.\ Stern et al.\ 1995). These models
could explain the steady Comptonized emission by basically adjusting one 
parameter,  the ratio of heating to cooling compactnesses. 
When supplemented with the idea of plasma outflow (Beloborodov 1999a,b) 
or a hot  ionized skin (Nayakshin \& Dove 2001), 
they cloud explain the diversity of spectra and correlations between 
spectral parameters.
However, in order to explain the complex time
variability, a number of additional features had to be added:
correlations between flares (avalanches) to explain the PSD
and spectral evolution during a flare as a result of, e.g.,
rising the active region above the disc (PF99).
Similar developments can be expected in the hot flow scenario.

Precise geometry of the X--ray emitting structures in the hot flow
is obviously uncertain, but certain constraints can be obtained.
For example, compact active regions (size much smaller than the
inner disc radius) moving through hot plasma totally 
devoid of optically thick plasma (i.e.\ the cold disc sharply disrupted
at certain radius), would intercept so few soft photons at $r<\rtr$
that the continuum slope would be much harder than observed. 
One possibility would then be that the active regions are compact
but the disappearance of the optically thick phase is gradual, moving
from $\rtr$ inwards. This could mean that small optically thick clouds form 
inside the hot plasma at $r<\rtr$ and their filling factor gradually
decreases inwards from $\approx 1$ to 0. Detailed simulations would
be necessary to predict spectra from such a configuration (e.g.\
Malzac \& Celotti 2002). At a rather speculative level one might
point out two possibilities of formation of such compact traveling 
active regions. The accreting plasma might form
multiple shocks where the dissipation of energy would preferentially 
take place, or the active regions might have something to do with
the dissipation of magnetic energy inside the hot flow, which dissipation
is necessary to maintain the equipartition between the magnetic and
thermal energy (Bisnovatyi-Kogan \& Lovelace 2000).

Another possibility for the geometry is to assume that the disc is rather
 sharply truncated
but the X--ray emission comes from global coherent perturbations propagating 
through the entire volume of the hot plasma. Then the decrease of 
$\lsoft/\lh$ would roughly correspond to the decrease  of the solid angle
subtended by the cold disc, as the perturbation approaches the center.

Temporal profiles of the flares considered in this paper were rather
simple, therefore to produce a broad band PSD a distribution of parameters 
(e.g.\ flare timescale) was required.
One might envision more complex profiles based on, e.g., 
recent considerations of Maccarone \& Coppi (2002). They suggest correlations
between launch times of short flares ($\trise\approx 0.5$ sec), 
such that the short flares
fill in a specific temporal envelope function. We note that their flares
do not explain the whole range of time-scales, since the PSD above
$\sim 1$ Hz is under-predicted. Even shorter flares would be necessary to
generate enough power at sub-second time-scales, perhaps filling in
the short flares considered by Maccarone \& Coppi (2002) as envelopes. 
The result might then be equivalent to a hierarchical scenario proposed by
 Uttley \& McHardy (2001),
where a long time-scales perturbation propagate from outside, breaking
down into smaller, faster structures in a fractal-like manner.

\section*{Acknowledgments} 
 
This work  was partly supported by grant no.\  2P03D01718 
of the Polish State Committee for Scientific Research (KBN).

{}


\end{document}